\documentclass[%
 aip,
 amsmath,amssymb,
 reprint,%
]{revtex4-1}

\usepackage{graphicx}% Include figure files
\usepackage{dcolumn}% Align table columns on decimal point
\usepackage{bm}% bold math
\usepackage[utf8]{inputenc}
\usepackage[T1]{fontenc}
\usepackage{mathptmx}
\usepackage{etoolbox}
\usepackage{color}

\makeatletter
\def\@email#1#2{%
 \endgroup
 \patchcmd{\titleblock@produce}
  {\frontmatter@RRAPformat}
  {\frontmatter@RRAPformat{\produce@RRAP{*#1\href{mailto:#2}{#2}}}\frontmatter@RRAPformat}
  {}{}
}%
\makeatother

\begin{document}

\preprint{AIP/123-QED}

\title{Emerging double power-law through dynamical complex networks in motility-induced phase separation in Active Brownian Particles}

\author{Italo Salas}
\email{italo.salas@ug.uchile.cl}
\affiliation{Departamento de F\'isica, Facultad de Ciencias, Universidad de Chile, Santiago Chile.}

\author{Francisca Guzm\'an-Lastra}
\email{fguzman@uchile.cl}
\affiliation{Departamento de F\'isica, Facultad de Ciencias, Universidad de Chile, Santiago Chile.}

\author{Denisse Past\'en} 
\email{denisse.pasten.g@gmail.com}
\affiliation{Departamento de F\'isica, Facultad de Ciencias, Universidad de Chile, Santiago Chile.}

\author{Ariel Norambuena}
 \email{ariel.norambuena@umayor.cl}
\affiliation{Centro Multidisciplinario de Fisica, Universidad Mayor, Camino la Piramide 5750, Huechuraba, Santiago, Chile.}%

\date{\today}% It is always \today, today,

\begin{abstract}
We investigate the behavior of active Brownian particles (ABP) within a temporal complex network framework approach. We focused on the node degree distribution, average path length, and average clustering coefficient across the P\'eclet number and packing fraction region. In the single phase or gas region, particle interactions mirror a random graph, and the average number of unique interactions display a non-monotonic behavior with the packing fraction. As we ventured toward the Motility-Phase Separation (MIPS) frontier, a hybrid distribution with a time-evolving pattern emerged, combining Gaussian and power law components. Moreover, we discovered a double power-law distribution in the phase-separated region, with two characteristic slopes symbolizing the emerging gas and solid regions. Our approach involved various numerical and theoretical analyses to capture the role of the packing fraction and Péclet number in the topological properties of the dynamical complex network that arises during the ABP dynamics. These findings provide a robust understanding of a phase transition between two states and how this transition manifests as a random to scale-free behavior.
\end{abstract}

\maketitle

\section{Introduction}

Active Brownian Particles (ABP) serve as a valuable framework for investigating the non-equilibrium dynamics and statistical properties of self-propelled agents, whether they are living organisms or artificial entities \cite{romanczuk2012active,bechinger2016active}. In contrast to passive Brownian particles studied in random diffusive systems, ABP offers a new dimension of complexity due to the inherent particle self-propulsion mechanisms. In particular, these active particles undergo a distinctive phase transition known as Motility-Induced Phase Separation (MIPS) \cite{fily2012athermal, redner2013structure,stenhammar2014phase,theers2018clustering,van2019interrupted}. This phase transition is related to the spontaneous separation between dilute (free-particles) and dense phases (cluster formation) \cite{stenhammar2014phase,de2023sequential,de2021active}. This phase separation is a fascinating emergent phenomenon where particles collapse into clusters although natural repulsion forces are present.
The exploration of ABP not only deepens into the understanding of self-organization and emergent phenomena but also opens up promising avenues for harnessing their unique properties in various applications and technological advancements \cite{gompper20202020} and building a bridge between physical models and biological systems.

The complexity of ABP dynamics and its statistical properties offer novel opportunities for deeper complex network analysis to better understand the role of particle connections during the dynamics. The particle interaction or contact dynamic between particles within this system is still a significant topic of interest, where the effects of particle shape \cite{theers2018clustering}, inertia \cite{caprini2022role}, hydrodynamic interactions \cite{zottl2023modeling,theers2018clustering,tan2022odd,petroff2015fast} and quorum sensing \cite{ridgway2023motility} among others has been studied numerically and analytically in the context of motility-induced clustering and motility-induced separation. 
In recent years agent-based-models such as ABP have been used to construct from particle direct physical contact the contagion dynamics on the top of compartmental ecological models \cite{norambuena2020understanding,de2023sequential,forgacs2022using}, or from more complicated pathways where temporal networks has been used to track burstiness or clusterization in the system \cite{richardson2015beyond,gorochowski2017behaviour,zhong2023burstiness,bhaskar2021topological}. 
In recent work, McDermott et al. unveil the time evolution hidden in the clustering effect during MIPS using machine learning \cite{mcdermott2023characterizing}, where they demonstrate different regimes close the phase separation.

In the last years, studies conducted by complex networks have been applied to many natural systems such as social interactions \cite{newman2001,Kertesz,gonzalez2008understanding}, seismology \cite{abe2006,telesca2012,pasten2016,pasten2018}, space plasmas \cite{suyal2014visibility,mohammadi2021complex}, or biological interactions\cite{thiery,barabasi2011,scabini,riley2007large,stockmaier2021infectious}, among others. The formalism of complex networks can show a non-trivial behavior in systems due to their geometrical and topological foundations \cite{newman2003,albert2002statistical}. Throughout a complex network analysis, we can characterize and describe a physical system based on its topological properties. Thus, we can classify a complex network based on its structural organization of connections between nodes such as a random network \cite{erdos1960}, a scale-free network \cite{barabasi1999}, or a small-world network \cite{watts1998}, for example. Due to the random encounters between ABP, one can construct a temporally averaged network that allows us to analyze the dynamics in terms of the topology of such connections. This new approach may motivate new questions or characterize the phase transition (MIPS) using the language of complex networks.

In this work, we carry out a new manner of analyzing ABP using complex networks: we build a time-averaged network with the time evolution of a system of ABP, and we follow the changes in the topological features of the complex network through the degree distribution ($P(k)$), the clustering coefficient ($\bar{C}$) and the average path length ($l_G$). Our approach suggests that a random graph can characterize the dilute phase while a double power law distribution describes the MIPS region.

\section{Active Brownian Particles (ABP) and motility-phase-separation}

\begin{figure}[ht]
    \centering 
    \includegraphics[width=\columnwidth ]{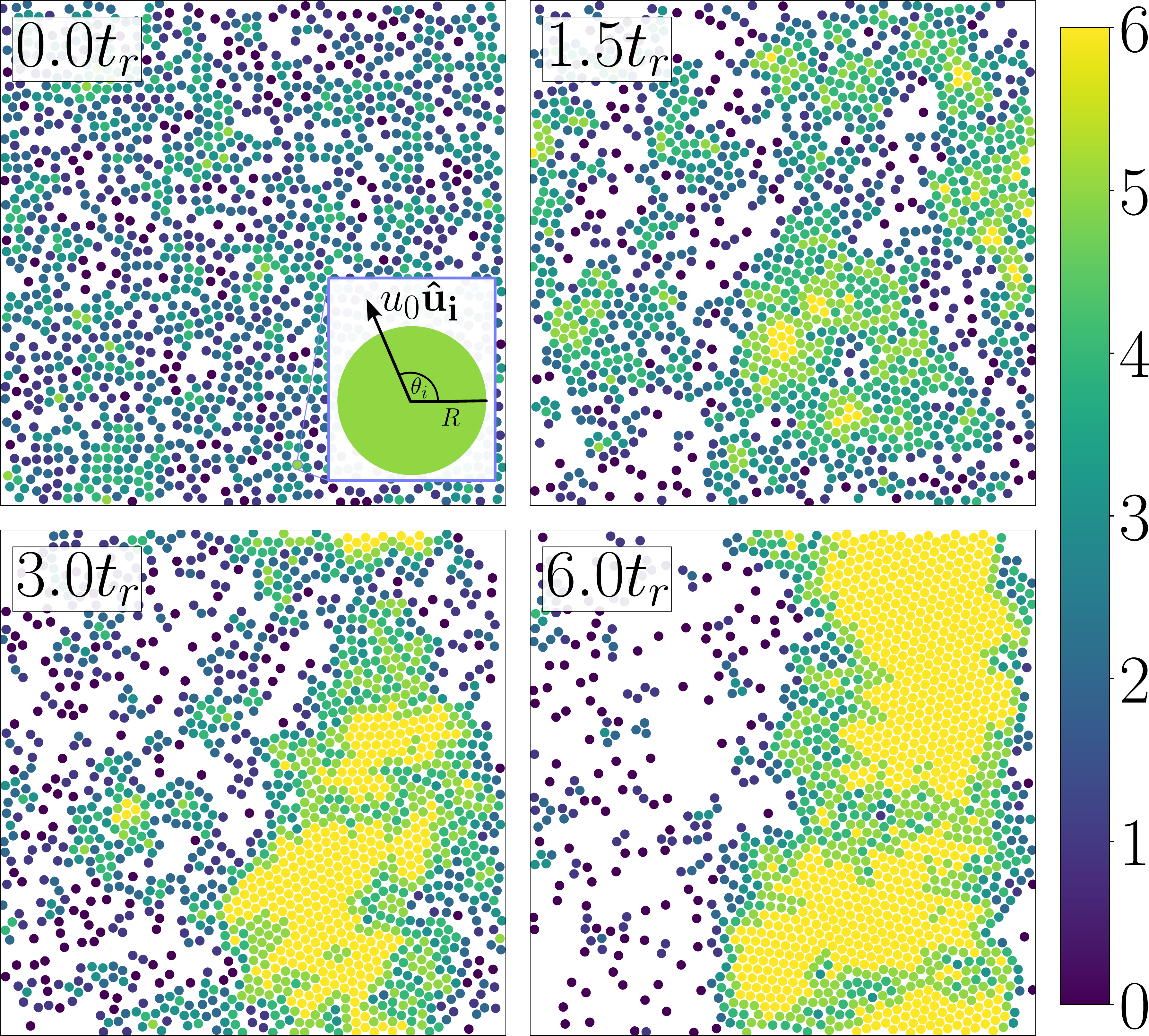}	
    \caption{Time evolution for a system of active Brownian particles exhibiting Motility-induced phase separation:  In the top-left inset, a single ABP is shown with orientation vector $\hat u_i$ and radius $R$. The time evolution is shown at four different times, showing a transition from a single phase in $t=0.0 t_r$ and two coexisting phases in $t=6.0 t_r$, at $\text{Pe} = 140$ and $\phi = 0.5$, with $ L = 100$ and $t_r = 1/D_R$.} 
    \label{fig:1}%
\end{figure}

We consider $N$ self-propelled disk-like particles with position vectors $\mathbf{r}_i(t)=[x_i(t),y_i(t)]$ (center of the disk) and orientation vectors $\hat{\mathbf{u}}_i(t)=
[\cos\theta_i(t),\sin\theta_i(t)]$ that obey overdamped Langevin dynamics in a flat 2D surface,
\begin{eqnarray}
\dot{\mathbf{r}}_i&=&u_{0}\hat{\mathbf{u}}_i(t)+(1/\gamma_T)(\boldsymbol{\xi}_{i,T}(t)-\nabla_{\mathbf{r}_i}U),  \label{eq1}\\
    \dot{\hat{\mathbf{u}}}_i&=&(1/\gamma_R)\boldsymbol{\xi}_{i,R}(t)\times\hat{\mathbf{u}}_i\label{eq2}
\end{eqnarray}

where $\gamma_T,\gamma_R$ are the Stokes friction coefficients associated with translation (T) and rotation (R) in the $x-y$ plane, respectively, $u_0$ 
represents the self-propelled velocity with constant magnitude see Fig.~\ref{fig:1}. Also, in Eqs.\eqref{eq1} 
and \eqref{eq2}, $\boldsymbol{\xi}_{T}(t)$ and $\boldsymbol{\xi}_{R}(t)$ are, respectively, the translational and rotational, Gaussian, white noises with zero mean, $\langle\boldsymbol{\xi}_{T}
(t)\rangle=\langle\boldsymbol{\xi}_{R}(t)\rangle=\mathbf{0}$, and time correlation $\langle\boldsymbol{\xi}_{T}(t_1)\boldsymbol{\xi}_{T}(t_2)\rangle=D_T\delta(t_1-t_2)$ and  $\langle
\boldsymbol{\xi}_{R}(t_1)\boldsymbol{\xi}_{R}(t_2)\rangle=D_R\delta(t_1-t_2)$, where we can define $D_T$ and $D_R$ as the 
translation and rotation diffusion parameters. Finally, in Eq.\eqref{eq1}, $U$ corresponds to the summation over steric interactions between particles at time $t$,

\begin{equation}
    U=\sum_{i\neq j}U_{\text{WCA}}(\mathbf{r}_i,\mathbf{r}_j)\label{eq.3}
\end{equation}

with $\mathbf{r}_j$ is the position of the $j-$th particle, and $U_{\text{WCA}}$ is the Weeks-Chandler-Andersen (WCA) pair potential 
 
\begin{equation}\label{eq.4}
U_{\text{WCA}}(\mathbf{r}_i,\mathbf{r}_j) =  \left\{\begin{array}{cc}
     \displaystyle{4 \varepsilon\left[\left(\frac{\sigma}{r_{i j}}\right)^{12}-\left(\frac{\sigma}{r_{ij}}\right)^{6}\right]}    &   r_{i j} \leq C_{r}\\
     0    & \mbox{otherwise} 
    \end{array} \right.
\end{equation}

with $r_{ij} = |\mathbf{r}_i-\mathbf{r}_j|$ is the distance between particles, $\sigma$ the particle diameter, $C_r = 2^{1/6} \sigma \approx 1.1225 \sigma$ is defined here as the contact radius, and  $\varepsilon$ is the energy interaction strength between particles. We set $\varepsilon = 3$ unless we stated otherwise during our simulations.

In this study, we conducted numerical simulations involving a collection of ABP moving within a square of area $L\times L$. Here, it is important to introduce the two dimensionless numbers to characterize different regimes. First, we introduce the packing fraction as $\phi=N\pi(\sigma/2)^2/L^2$, which relates the ratio between the area covered by the $N$ Active Brownian Particles with radius $\sigma/2$ with respect to the total available area. When the packing fraction increases, the high density allows clusters to be formed. In addition, these ABP systems exhibit a characteristic persistent motion reminiscent of a run-and-tumble state observed in self-phoretic systems and bacterial suspensions \cite{cates2015motility, buttinoni2013dynamical,petroff2015fast,van2019interrupted}.
The persistence time is defined as $t_p=\sigma/u_0$, while the memory kernel associated with tumble states is characterized by $t_r=1/D_R$. Since ABP can explore the available space through diffusion, it is natural to define $t_r$ as the relevant time scale of the problem. Second, we introduce the P\'eclet number $\text{Pe} = t_r/t_p = u_0 t_r/\sigma$, which relates the ratio between diffusion and persistent times in transport phenomena. When the P\'eclet number is large, we have that ABP are persistent compared to diffusion. The latter is relevant for cluster formation since the minimal cluster for disk-shaped particles is formed with three persistent disks forming an equilateral triangle.

The interplay between steric interactions causing the blocking of persistent motion and an increased probability of collision due to elevated concentrations $\phi$ and particle activities $u_0$ leads to a fascinating phenomenon known as motility-induced phase separation (MIPS) \cite{theers2018clustering, stenhammar2014phase}. The system starts with a homogeneous particle distribution and orientations at $t=0$ exhibiting a single-phase behavior (or fluid-like state \cite{su2023motility}), which physically describes ABP randomly moving without particle agglomeration. However, if the packing fraction and P\'eclet number are larger than some critical values, the system evolves into a two-phase state represented by a solid and gas phase, as illustrated in Fig~\ref{fig:1}. In this phase, particles agglomerate, and cluster formation becomes an emergent phenomenon.

An analytical alternative to represent this phase separation is using the separation kinetic function $f_c$, which reproduces the phase separation map in terms of the packing fraction, the P\'eclet number, and the average total number of particles lost per escape event $\kappa$~\cite{redner2013structure}

\begin{equation}
    f_c=\frac{4\phi\text{Pe}-3\pi^2\kappa}{4\phi\text{Pe}-6\sqrt{3}\pi\kappa\phi}
    \label{eq:fc}
\end{equation}

\begin{figure}[ht]
    \centering 
    \includegraphics[width=\columnwidth]{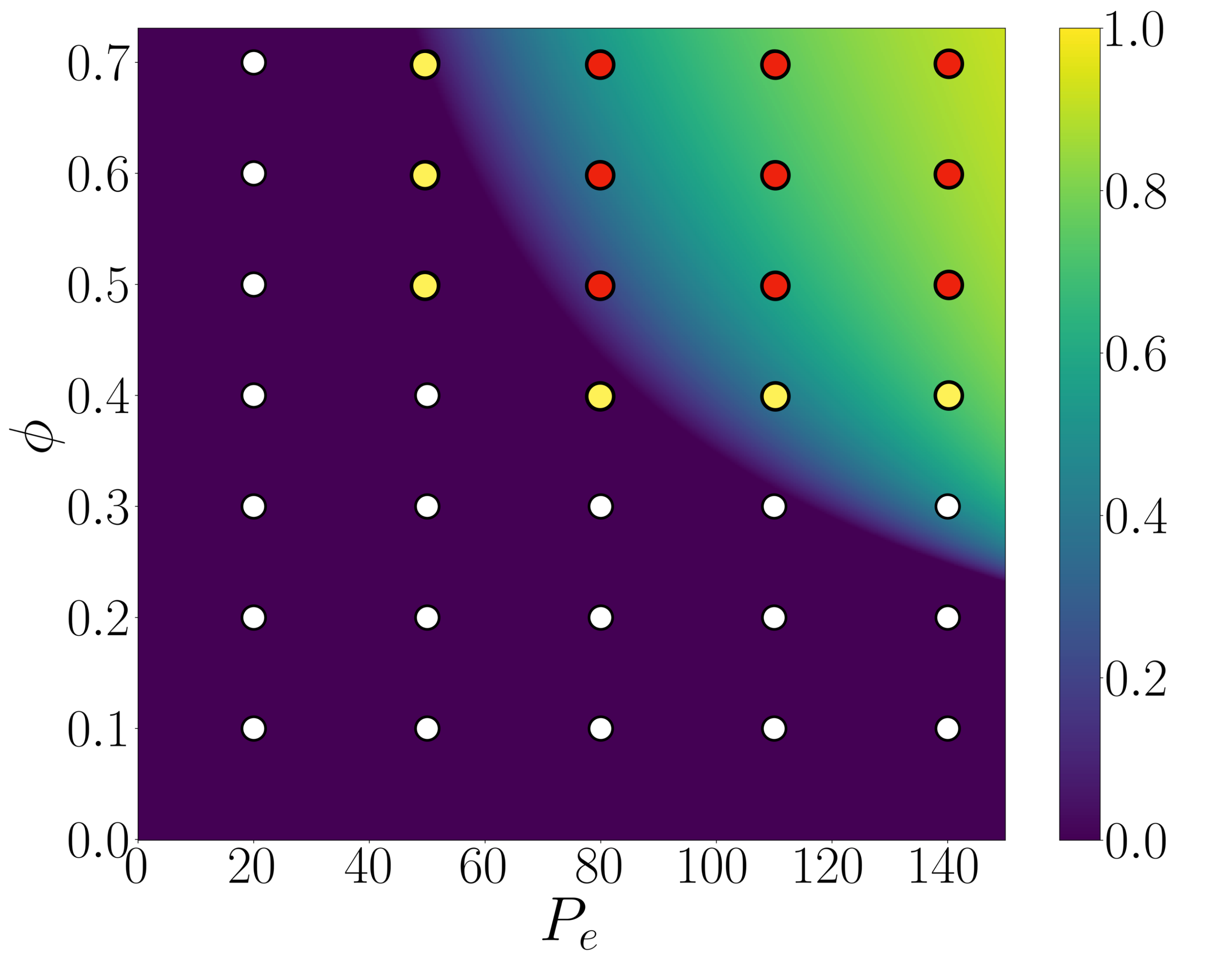}
    \caption{Analytical representation for the phase diagram in the $\text{Pe}-\phi$ plane plotted from the analytical model using Eq.~\eqref{eq:fc}, with $\kappa=4.7$ \cite{redner2013structure}. The white dots represent the pairs $(\text{Pe},\phi)$ parameters studied across this work. The different colors represent the observed networks. White dots represent the single phase where a random graph is observed. The yellow dots represent the hybrid phase, and the red dots represent the two-phase region where a double power-law distribution is recovered. } 
    \label{fig:2}%
\end{figure}

In Fig.~\ref{fig:2}, we analytically reproduced the phase diagram in terms of the packing fraction and the P\'eclet number using $\kappa=4.7$ as a fit parameter \cite{redner2013structure}. The pairs $(\text{Pe}, \phi)$ studied in this work are represented by dots in Fig.~\ref{fig:2}. For $f_c = 0$, we can say that the system is in the single-phase (white dots), while for $f_c \neq 0 $, the system undergoes a phase transition to the two-phase state (red dots). The system is considered steady when the phase separation shows a bimodal probability distribution for different local packing fractions \cite{redner2013structure,stenhammar2014phase}. Despite extensive studies on phase separation in various contexts, machine learning has recently investigated the time evolution during phase separation, revealing different transition regimes \cite{mcdermott2023characterizing}. In this work, we aim to study the time evolution by generating a temporal averaged network during the phase transition.

\subsection{Simulation methods and parameters}

We employed standard Brownian dynamics methods to numerically solve the governing Eqs.\eqref{eq1} and \eqref{eq2} for $N \in [955,1274,1592,1910,2229]$ particles within a simulation box with a length of $L=100$ and periodic boundary conditions. We ran $20$ ensembles for each set of parameters, initiating simulations with random initial particle positions and orientations.

The relevant length scale in the system is the particle radius $R$, serving 
as the unit of length. Additionally, we adopted $t_r$ as the unit of time. 
The equations of motion, Eqs.\eqref{eq1} and \eqref{eq2}, were integrated 
with a time step of $\Delta t=3\times10^{-5}t_r$, and simulation times were set to $T = 30 t_r$. Two dimensionless values parameterize the system: the packing fraction $\phi$ and the P\'eclet number, which are varied in the intervals $\phi\in [0.3-0.7]$ and $\text{Pe}\in [80,140]$, respectively (see Fig.~\ref{fig:2}). We maintain a constant rotational diffusion coefficient $D_R=0.15$, translation diffusion coefficient $D_T=0.2$, the diameter $\sigma=2R$ with $R=1$, and the Stokes translational and rotational friction coefficients are $\gamma_T=6\pi$ and $\gamma_R=8\pi$ respectively.

\section{Complex Networks for ABP}

A network is defined as a group of nodes connected between them by edges. The connections between nodes could be directed or undirected. In this case, the nodes are determined by each ABP, and the interaction between particles defines the connections: two nodes (particles) are connected when their centers are inside the interaction region defined by the contact radius $C_r$ (see WCA potential in Eq.~\eqref{eq.4}). We build a time network by following the dynamics of the particle connections in a specific time window. The network is built considering a fixed number of nodes $N$ and an increasing number of connections.

\begin{figure}[ht]
    \centering 
    \includegraphics[width= 1 \columnwidth]{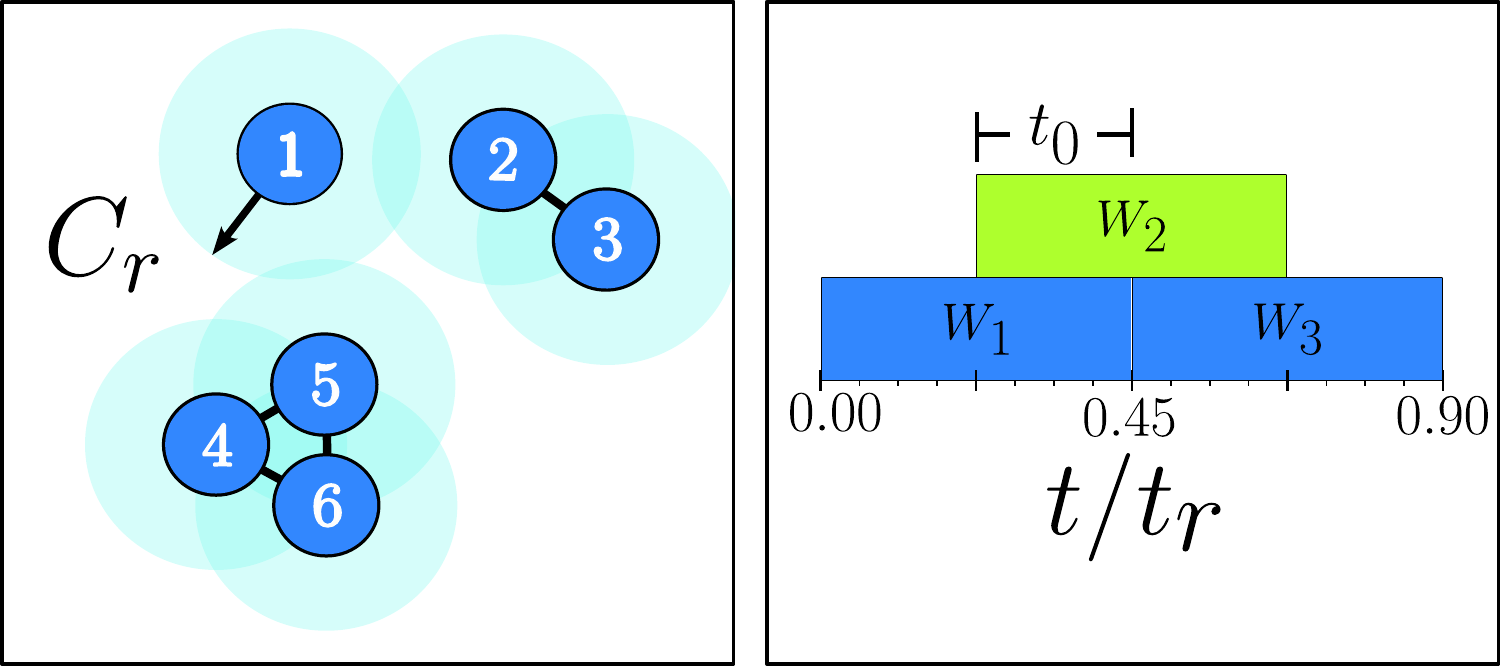}
    \caption{From left to right: Scheme for particle encounters when a pair interaction occurs. The blurred area represents the interaction region with radius $C_r$. Particles (nodes) generate a link if their centers enter each other's interaction region, as shown in the middle panel. A representation of the time evolution for the network formation, with time windows of length $W=0.45 t_r$ and overlap $t_0=0.225 t_r$ between two consecutive windows.} 
\label{fig:3}%
\end{figure}

To have enough statistics, we define a window with sixty data points, and we count the connections between nodes each time; if two nodes are connected only once in a one-time step or if two nodes are connected along the sixty steps, the value of the connections in the adjacency matrix is one in that time window. We move the window to the next time window's center and repeat the same procedure, as shown schematically in Fig.~\ref{fig:3}.

\subsection{Adjacency matrix}

In order to analyze the system, we first create the 
adjacency matrix $A$, which is used to represent the network. Using the distance $r_{ij} = |\mathbf{r}_i - \mathbf{r}_j|$ between the centers of two particles, we can establish the value of $A_{ij}$ 
which represents the connections between these particles. If $r_{ij} \leq C_r$, we set $A_{ij}=A_{ji} 
= 1$; otherwise, $A_{ij} = 0$. If $A_{ij}=A_{ji} 
= 1$ it means that two nodes (particles) are connected between them, and the complex network built is undirected. Within time windows of length $W = 0.45 t_r$ and using an overlapping time $t_0 = 0.225 t_r$, a link is established only upon the first interaction between two particles. The latter means that each network accounts for unique particle encounters within the time window.

From the adjacency matrix, we can compute all the necessary properties of the network. As we know, the study of complex networks can shed light on the global behavior of a system composed of small and separated elements. We will focus on calculating the degree distribution, network average clustering coefficient, and average path length. These metrics give good insights into describing, characterizing, and understanding the time evolution of the structure of a complex network.   

\subsection{Degree distribution}

As a first manner to characterize these complex networks, we compute the degree of the nodes. We define the degree $deg(p_i)$ of each  particle $p_i$ as the number of connections of a node, and it can be computed by summing over the corresponding column or row of the adjacency matrix:

\begin{equation}
    deg(p_i) = \sum_{j=1}^{N} A_{ij} = \sum_{j=1}^{N} A_{ji},
    \label{eq5}
\end{equation}

and the degree distribution is defined by 

\begin{equation}
    P(k) = \frac{n_k}{N},
    \label{eq6}
\end{equation}

which is directly obtained from simulations. Here $n_k$ is the total number of particles with degree $k$, and $N$ is the number of particles in the system. Depending on the topology of the complex network, the degree distribution could show different behaviors~\cite{Erdos,barabasi1999,albert2002statistical,newman2003}: it could be binomial (random graphs) or power-law (scale-free). Real networks usually show a scale-free behavior. To explain this behavior of a complex network built with the World Wide Web, Réka Albert and Albert Barabási~\cite{barabasi1999,albert2002statistical,barabasi2011} used a model for random networks given a preferential probability of attachment for the new connections.  
In this work, we shall discuss the role of the Gaussian distribution and double power law models in the context of MIPS to understand the global structure of these complex networks. Also, the degree distribution has been shown to be useful to identify phase transitions in real networks~\cite{pasten2016}. 

\subsection{Network average clustering coefficient}

Another characteristic of real networks is the small-world phenomenon. Watts and Strogatz~\cite{watts1998} show that random networks do not have this behavior in which two nodes of the complex network are connected between them through small steps. In order to give a measure of the small world behavior, we can define a global clustering coefficient (for all the nodes of the network) and a local clustering coefficient (for each node). The values of the clustering coefficient allow us to identify these small-world properties. Both random networks and ordered lattices do not show a small-world behavior. The clustering coefficient~\cite{Newman_2000,newman2003} $C_i$ of each particle is a measure of the tendency to form clusters of this node, and it is defined by

\begin{equation}
    C_i = \frac{1}{k_i(k_i-1)}\sum_{j,k} A_{ij}A_{jk}A_{ki}, 
\end{equation}

where $A$ is the adjacency matrix, and $k_i$ is the degree of the particle. From this, we can calculate the network average clustering coefficient~\cite{watts1998} $\bar{C}$, which is the trend of the network to form clusters, and it is defined by

\begin{equation} 
    \bar{C} = \frac{1}{N} \sum_{i=1}^{N} C_i, \quad \quad 0 \leq \bar{C} \leq 1.
    \label{clustering}
\end{equation}

\subsection{Average path length}

To identify the small-world phenomenon, we need an additional measure: the distance between two nodes, called average path length~\cite{Newman_2000}. The small-world model shows an increase in the log scale of the average path length and the clustering properties at the same time. The average path length is the rate between the shortest number of steps between two nodes and the shortest paths between all the pairs of nodes, and it is defined by~\cite{Newman_2000},  

\begin{equation}
    l_G = \frac{1}{N(N - 1) } \sum_{i \neq j} d(v_i,v_j). 
    \label{lg}
\end{equation}

Here, $d(v_i,v_j)$ is the shortest distance between two vertices $v_i$ and $v_j$, where $v_{i,j}$ are nodes of the graph.
The small-world model shows a single length scale that depends on the probability of the connections between nodes~\cite{newman_watts1999}. 

\section{Results and Discussion}

We studied three different regimes of the ABP according to the phase diagram of Fig.~\ref{fig:2}: (i) one-phase (white circles), (ii) transient phase (yellow circles), and (iii) two-phase (red circles). As discussed in the following sections, these three regions will exhibit different topological properties.

\subsection{One-phase region: Random Graph}

In the one-phase region (violet in Fig.~\ref{fig:2}), we found that the random motion of ABP can be characterized by a Gaussian function for the degree distribution

\begin{eqnarray} \label{GuassianDistribution}
    P^{\rm G}(k) = {\frac{1}{ \sigma_{\rm st}  \sqrt{2\pi}}} \mbox{exp}\left[ - \frac{1}{2}\left( \frac{k-\mu}{\sigma_{\rm st}}\right)^2 \right],
\end{eqnarray}

where $\sigma_{\rm st}$ is the standard deviation, and $\mu$ is the average number of unique encounters of particles in a time window $W$. As we shall illustrate later, We have a dilute system for packing fraction numbers $\phi < \pi/(\pi+2^{13/6})$, which allows us to analyze the average behavior of the ABP system using the kinetic theory of gases. As explained in Ref.~\cite{norambuena2020understanding}, in an active media with 
$N$ moving particles in a dilute system, the mean free path is $\lambda = \sqrt{2} u_0(\phi) \tau_c$, where $\tau_c$ is the mean time between encounters and we use $u_0(\phi) = u_0(1- \chi \phi)$ as the reduced velocity~\cite{soto2024kinetic}. During a time interval $\tau_c$, we can establish the probability constraint $p_c = A_{\rm sw} /L^2 \approx 1$, where $A_{\rm sw} = N(2\lambda a + \pi a^2)$ 
is the total area swept area of the $N$ particles, and $L^2$ is the available 
area. From this argument, it follows that $\lambda =  (\pi/4) \phi^{-1}(1-\phi) 
\sigma = \sqrt{2} u_0(\phi) \tau_c$. We remark that this expression for the mean free path is only valid for $\lambda > C_r$ (critical radius), which implies that $\phi < \phi_c = \pi/(\pi + 2^{13/6}) \approx 0.41$. As we shall discuss later, the critical packing fraction $\phi_c$ sets a critical density for the dilute phase (one-phase). Then, the average number of encounters can be estimated as $\mu = P(\lambda) W/\tau_c$, where $P(\lambda)$ is the encounter probability between agents at the region $\sigma<r<\lambda$. This probability can be estimated as $P(\lambda)= \int_{\sigma}^{\lambda}p(l)dl$, where $p(l) = \lambda \mbox{exp}(-(l-\sigma)/\lambda)$ is the probability distribution of the mean free path in a gas condition~\cite{Paik_2014}. Then, we obtain

\begin{equation}\label{mu}
    \mu(\phi,\text{Pe}) = \Lambda(\text{Pe})  \left({\phi \over 1-\phi} \right)(1-\chi \phi) P(\lambda),
\end{equation}

\begin{figure*}[ht!]
    \centering 
    \includegraphics[width=\textwidth]{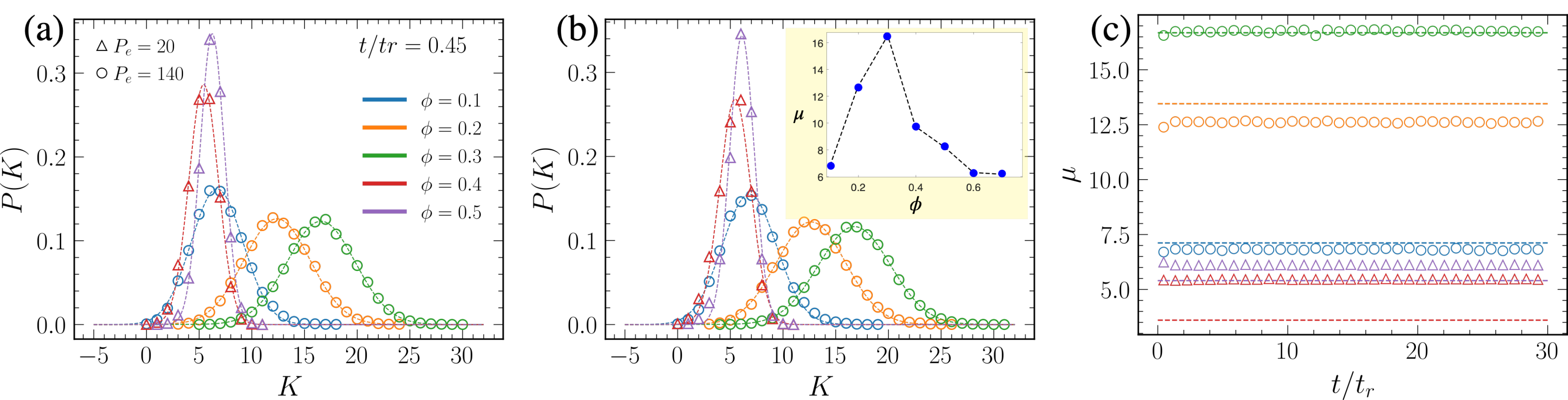}
    \caption{(a), (b) Time evolution for the degree distribution $P(k)$ for systems in the one-phase region at $\text{Pe} = 20$ (triangles) and $ \text{Pe} = 140$ (circles) at time $t = 0.45 t_r$ and at  $t = 30.0t_r$ for different packing fractions $\phi$. The distributions show a Gaussian tendency; each system exhibits distinct mean and standard deviation values. Note that all systems maintain a Gaussian distribution. The inset shows the average number of unique encounters between particles $\mu$, for different \text{Pe} and $\phi$, extracted from (c). (c) The average number of unique encounters of particles $\mu$ for different P\'eclet numbers in the single-phase, dashed lines show the analytical value following Eq.~\eqref{mu}.} 
    \label{fig:4}%
\end{figure*}

where $\Lambda(\text{Pe}) = 4 \sqrt{2} W D_R \text{Pe}/\pi$, $P(\lambda) = 1-e^{-(\lambda- \sigma)/\lambda}$, $\lambda = \lambda(\phi)$, $W = 0.45 t_r$, $D_R = 0.15$, and we use $\chi = 0.2$ to have a better fit. In Fig.~\ref{fig:4} (a), (b), we show the time evolution for the Gaussian distributions of $P(k)$ for different packing fractions and P\'eclet numbers. Using our model given in Eq.~\eqref{mu}, we obtain $\mu(\phi=0.1,\text{Pe}=140) \approx 7.118$, $\mu(\phi=0.2,\text{Pe}=140) \approx 13.456$, and $\mu(\phi=0.3,\text{Pe}=140) \approx 16.686$
whose values are close to those obtained by our simulations ($6.804$, $12.659$, $16.773$), see Fig.~\ref{fig:4} (c). This simple model can give good estimations of the central value of the Gaussian 
distribution $P^{\rm G}(k)$ given in Eq.~\eqref{GuassianDistribution} in the regime $\phi < \phi_c=0.41$.
The inset in Fig.~\ref{fig:4} (b) resumes the obtained values for $\mu$ in our simulations. Showing a non-monotonic behavior, where $\mu$ increases with $\phi$ for dilute concentrations and $\mu$ decreases for denser systems.

\subsection{Transient Region: Hybrid phase}

\begin{figure}[ht!]
    \centering 
    \includegraphics[width=\columnwidth]{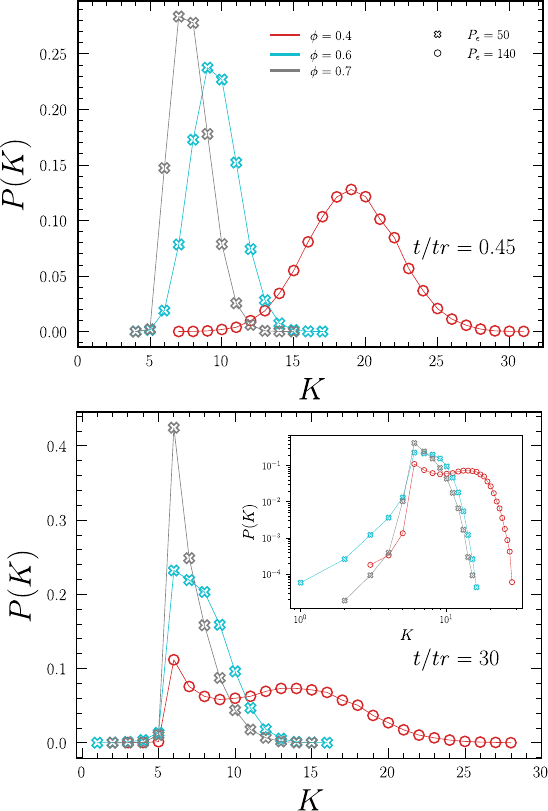}
    \caption{Time evolution for the degree distribution at the hybrid phase. At the initial time $t = 0.45 t_r$, a Gaussian distribution is observed for the different systems; meanwhile, a hybrid distribution is observed at time $t = 30.0t_r$. The inset shows the same data plotted in $\log{} - \log{}$ scale.} 
    \label{hybrid}
\end{figure}

We define the transient region as the intermediate phase between the one-phase and two-phase regions (blue in Fig.~\ref{fig:2}). Since it is an intermediate phase, it is expected to have a degree of distribution with a hybrid behavior. Physically and analyzing our numerical simulations, we note that in this regime, small clusters composed of three disks (triangular arrangement) appear and eventually break because of the persistent ABP motion. In Fig.~\ref{hybrid}, we observe the time evolution for the degree distribution at times $t=0.45t_r$ and $t=30t_r$. Interestingly, the initially random graph (Gaussian distribution) evolves to a hybrid distribution with two components (Gaussian and power law). This hybrid behavior tells us that the system is trying to evolve from a random to a scale-free graph, but the values for the packing fraction and P\'eclet are not large enough. Moreover, at $t=30t_r$, the system does not reach the steady state, which is not the case for the one-phase and two-phase regions. The latter tells us that this hybrid phase is more complex to analyze regarding its geometrical and topological properties since the system remains out of equilibrium.
In Fig.~\ref{fig:2}, we colored the data points where we observed this regime in yellow circles.

\subsection{Two-phase region: Double power law distribution}

\begin{figure*}[ht!]
\includegraphics[width=0.9\textwidth]{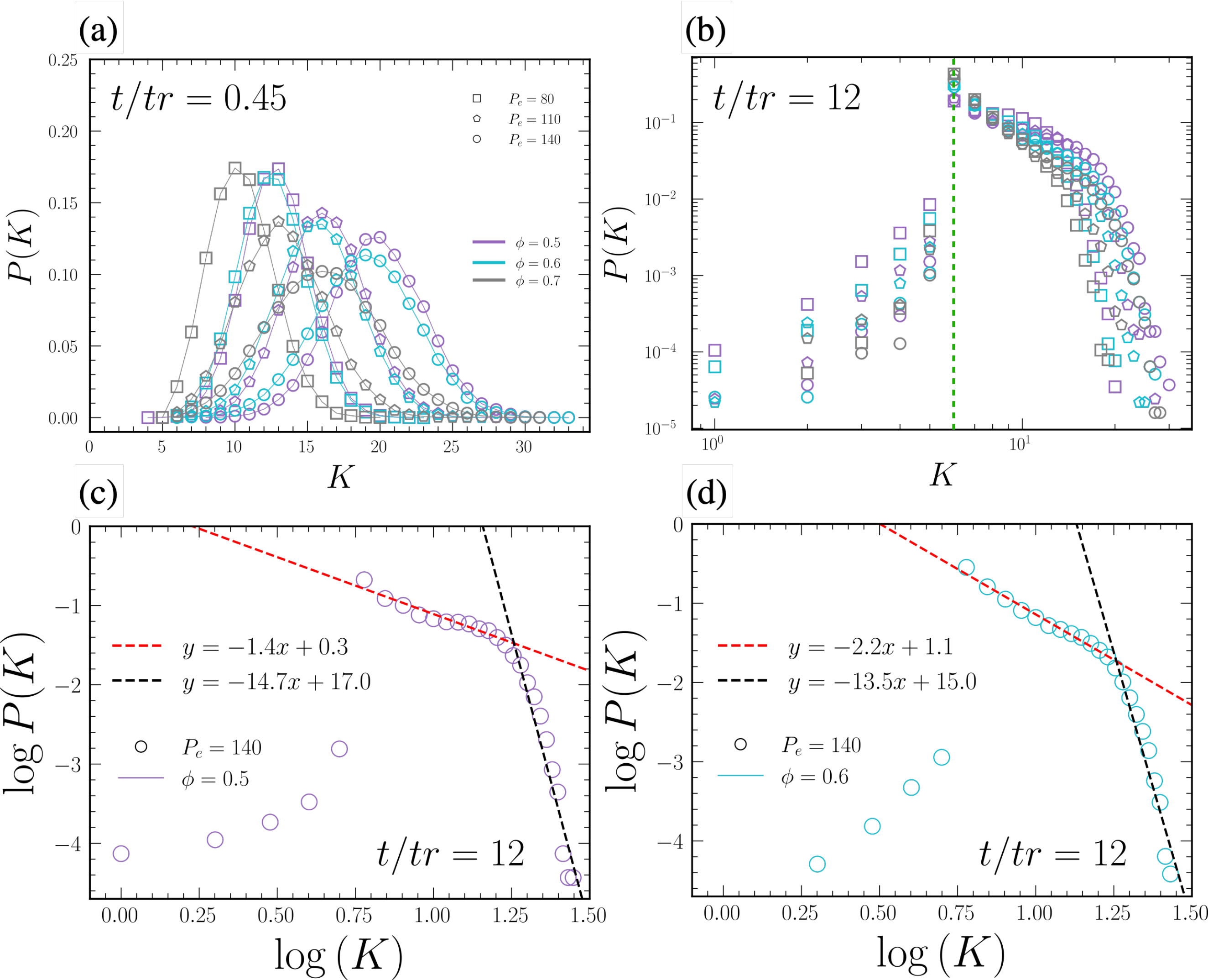}	
    \caption{Degree distribution for phase-separated systems. (a) $\log{} - \log{}$ plot shows a clear Gaussian distributions at time $t = 0.45 t_r$. (b) degree distribution at $t = 12.0 t_r$ shows an evolution in the distribution where the degree $k = 6$ for each curve is the one with the highest probability. (c) and (d) shows two distributions of figure (B) zoomed in at $\text{Pe}, \phi = (80,0.7) $ and  $\text{Pe}, \phi = (140,0.5)$ respectively, where the red and black dashed lines are the best linear fit for the first and second power law decay respectively for each distribution.} 
\label{fig:6}%
\end{figure*}

At more significant packing fractions and P\'eclet numbers, i.e., $\phi\geq 0.5$ and $\mbox{Pe} \geq 80$, the system enters the inner region of Fig.~\ref{fig:2}, where MIPS is manifested. From the point of view of the ABP, as time evolves, the initially random distribution (liquid-like state) evolves to a state where a gas-like and a solid-like phase cohabit in the system (green-yellow regions in Fig.~\ref{fig:2}). In terms of the complex network, the degree distribution of the system evolves from a Gaussian distribution at short times ($t=0.45 t_r$) towards a double power law distribution for larger times ($t\geq 11 t_r$); see Fig.~\ref{fig:6} (a),(b). Notably, when the degree distribution is plotted in the log-log scale, two slopes adequately describe the distribution $P(k)$, see Fig.~\ref{fig:6} (c), (d). This double power law distribution has the form

\begin{equation}
    P(k>K) \sim \left\{\begin{array}{cr}
        k^{-\gamma_1}, & k < k_{c}  \\
        k^{-\gamma_2}, & k \geq k_{c}
    \end{array} \right.,
    \label{2plaw}
\end{equation}

where $K = 6$, $k_c \in [13-17]$ is the critical degree and $(\gamma_1,\gamma_2)$ are the critical exponents. The value $K=6$ is connected with the 2D disk packing problem, where each disk has exactly six nearest neighbors in an ideal cluster and corresponds to the peak of $P(k)$, see Fig.~\ref{fig:6} (b). Moreover, the critical values for $k_c$ depend on the clustering coefficient and P\'eclet number. The first slope $\gamma_1$ (red dashed line of Fig.~\ref{fig:6} (c), (d)) changes from values close to $1.5$ at $\phi=0.5$ towards values close to $3.7$ at $\phi=0.7$. The second slope $\gamma_2$ (black dashed lines) varies from values close to $12$ to $17$. In summary, we obtain $\gamma_1 \approx (1.5-3.7)$ and $\gamma_2 \approx (12-17)$, and therefore $\gamma_1 < \gamma_2$ in our ABP model. In Table~\ref{tab:table1}, we compare our results with different systems that exhibit a double power law distribution.

\begin{figure}[ht!]
    \centering 
    \includegraphics[width=\columnwidth]{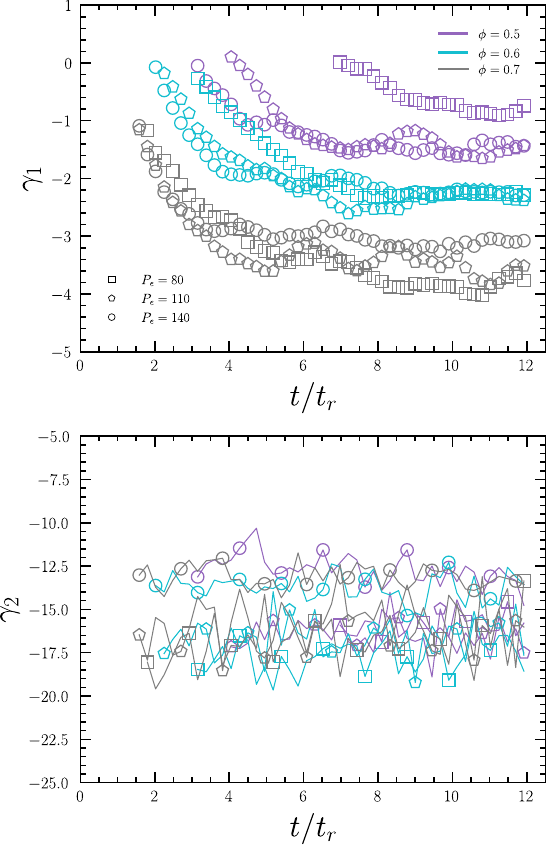}
    \caption{Time evolution for the slopes fitted for the power law distribution \eqref{2plaw}, in the phase-separated systems.} 
    \label{fig:7}
\end{figure}

Physically, the $\gamma_1$ exponent is related to the solid cluster appearance where the degree number has a peak at $k=K$ ($\mbox{log}_{10}(6) \approx 0.78$). To better understand this, we plot the time evolution in Fig.~\ref{fig:7}, where it is clear that the slope value and, therefore, the cluster solid-like region is determined strongly by the system packing fraction $\phi$. On the contrary, the $\gamma_2$ exponent is related to less probable encounters related to the gas region, and we will not delve into that aspect.

The temporal behavior of both curves is shown in Fig.~\ref{fig:7} ($\gamma_1$) and ($\gamma_2$). We note that $\gamma_1$ is more stable over time, and the packing fraction dominates its steady value in some inverse relation. On the contrary, $\gamma_2$ shows more fluctuations, which can be explained because the clustering formation is unstable for the values of $(\phi,\mbox{Pe})$.

\begin{widetext}

\begin{table}[h!]
  \begin{center}
    \caption{Comparison between critical exponents for the double power law distribution in different models. Here, we are using the acronyms CAN = Chinese Airline Network, ABP = Active Brownian Particles, and MIPS = Motolity-Induced Phase Separation.}
    \label{tab:table1}
    \begin{tabular}{c|c|c|c} 
      \textbf{System } & \textbf{Network } & $\gamma_1$  & $\gamma_2$ \\
      \textbf{or Data} & \textbf{(nodes, edges)} & approx.  & approx. \\
      \hline
      CAN Data~\cite{Li2004} &  (airports,flights) & $0.41-0.45$ & $4.16-4.54$ \\
      US Flight Network Data~\cite{Li-Ping:1393} & (airports,flights) & $0.56-0.57$ & $3.27-3.41$ \\
      Evolution of CAN~\cite{Han_2011} & (airports,flights) & $0.46-0.52$ & $2.1-2.7$ \\
      Evolution of CAN~\cite{ZHANG20103922} & (airports,flights) & $0.49$ & $2.63$ \\
      Language model~\cite{Dorogovtsev2001} & (words, connected words) & $3/2$ & $2$ \\
      Homer's Odyssey Data~\cite{PINTO20144019} & (words, connected words) & $0.76$ & $1.19$ \\
      ABP (our work) & (particles, interacting particles) & $1.5-3.7$ & $12-17$\\
    \end{tabular}
  \end{center}
\end{table}

\end{widetext}

\subsubsection{Role of the average clustering coefficient}

According to our simulations, particles agglomerate when $(\phi, \mbox{Pe})$ cross some critical region, and a natural tendency to form clusters appears. In the language of complex networks, we can elaborate on the connection between this phase separation using the tendency to form clusters from a topological point of view. We encoded the topological information into a dynamical adjacency matrix constructed upon unique encounters in a time interval. Then, we can use the average clustering coefficient presented in Eq.~\eqref{clustering} to analyze the global property of such connections to learn how to interpret the inherent hexagonal packing of the solid disks used in our ABP model. \par 

In the upper panel of Fig.~\ref{fig:8}, we show the temporal behavior of the average clustering coefficient $\bar{C}$ varying the packing fraction ($\phi$) and P\'eclet number (Pe) and covering different points of the phase diagram illustrated in Fig.~\ref{fig:2}. In all cases, the natural tendency for a fixed value of Pe is to have higher values of $\bar{C}$ as the packing fraction increases, where $0.1 \leq \bar{C} \leq 5.4$. Also, the persistent motion represented by the P\'eclet number shows that for $\phi \geq 0.5$, an increment of Pe increases does not affect $\bar{C}$ in the long-time behavior, see upper right panel of Fig.~\ref{fig:8}. \par

To have an intuition about the higher values of $\bar{C}$, we show in Fig.~\ref{fig:9} the average clustering coefficient when particles agglomerate, forming perfect hexagonal clusters. In this ideal cluster, the asymptotic behavior is to reach a value $\bar{C} =0.4$, which can be understood as follows. Let be $\bar{C}_{hex}(N)$ the average clustering coefficient for a hexagonal arrangement with $N$ particles. It is not difficult to show that for $N$ particles, the fractions of particles inside the borders are given by $f_{i} = (1+3N_r(N_r-1))/(1+3N_r(N_r+1))$, where $N_r = [(1+4/3(N-1))^{1/2}-1]/2$. Moreover, each particle inside the cluster has a clustering coefficient $C_i = 6/15$. For the particles at the border, we have particles in the corners with clustering coefficient $C_{1} = 2/3$, and for the rest of the particles, we have $C_2 = 1/2$. The fraction of particles at the corners is $f_c = 1/N_r$. Based on this, we found the analytical expression:

\begin{equation}
\bar{C}_{hex}(N) = \left[ f_{c}(N)C_1 + (1-f_{c}(N))C_2 \right](1-f_i(N)) + C_i f_i(N),
\end{equation}

which gives us the formula to calculate the average cluster coefficient for an ideal hexagonal cluster. This expression perfectly matches the numerical data points presented in Fig.~\ref{fig:9}. Most importantly, this tells us that as $N$ increases, the system stabilized around $\bar{C} = 0.4$ since $\lim_{N \rightarrow \infty}\bar{C}_{hex}(N) = 6/15 = 0.4$.

\subsubsection{Role of the average path length}

\begin{figure*}[ht!]
    \centering 
    \includegraphics[width=\textwidth]{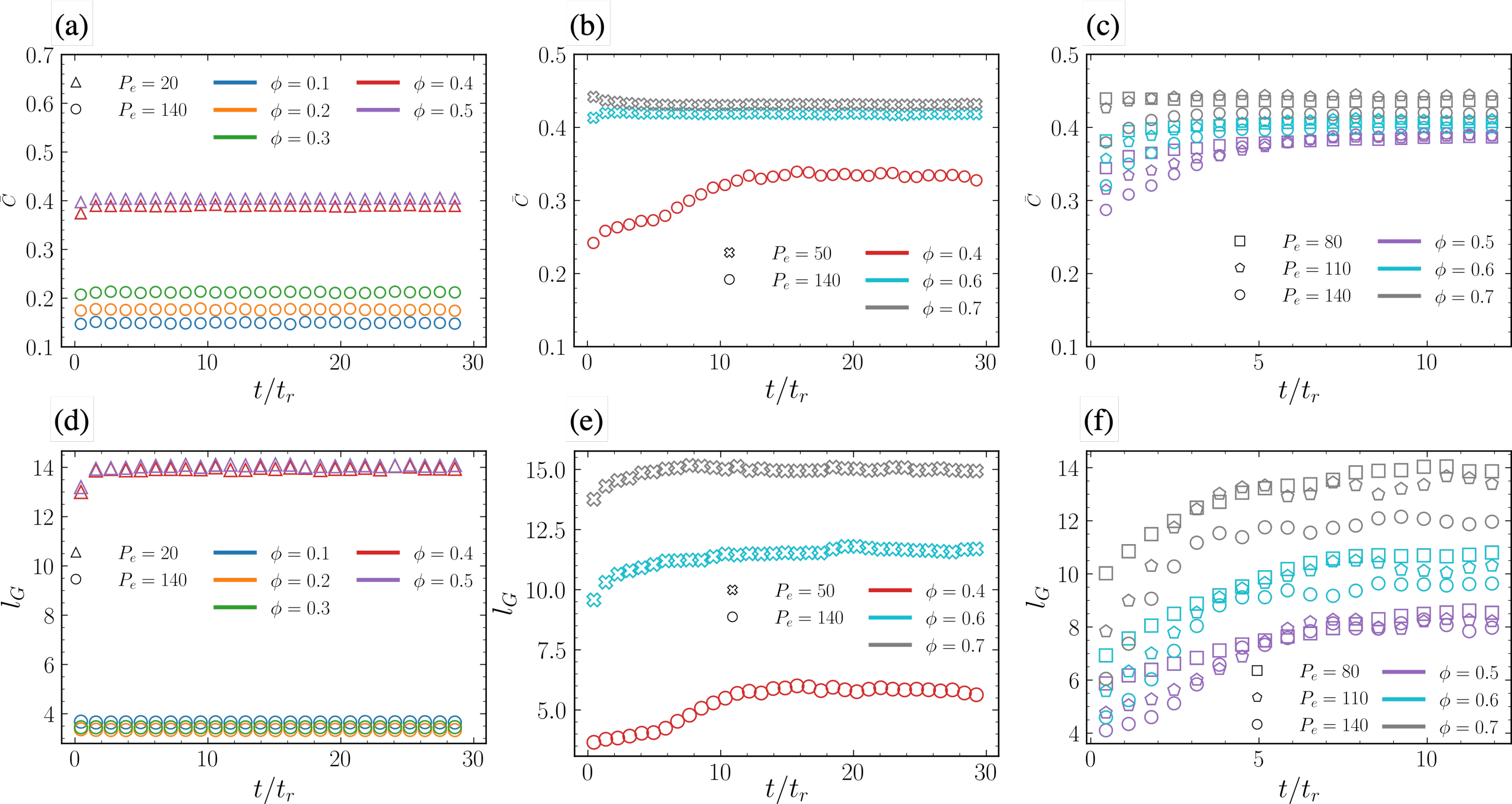}	
    \caption{Top: Network average clustering coefficient of identical circles in a hexagonal packing arrangement, using $C_r = 2^{1/6}\sigma$ ( black circles). $\bar{C}$ approaches asymptotically to $0.4$ (dashed red line), which is the value of the clustering coefficient of the center particle in the seven-particle arrangement (inset). Bottom: Average path length for the same hexagonal packing arrangement as a function of the packing fraction $\phi$ (black circles). We see that $l_G$ grows as $30 \sqrt{\phi}$ (red dashed line) and could represent an upper bound with respect to $\phi$ for clustered ABP systems.}
    \label{fig:8}
\end{figure*}

The average path length defined in Eq.~\eqref{lg} was measured through all regions, and their values are shown in the lower panel in  Fig.~\ref{fig:8}. We observe the time evolution for the network average path length in the case of a system with a single phase characterized by a Gaussian degree distribution $l_G$ has a constant value for different $\text{Pe}$ numbers and different packing fractions $\phi$.  
Meanwhile, on the transient and the two-phase region, the $l_G$ value increases slowly in time, reaching a constant value when $t/t_r\geq10$ for the hybrid phase and in $t/t_r\geq 8$ in the two-phase region, see Fig~\ref{fig:8} (e), (f) respectively. 
This time scale is similar to the cluster growth characteristic time reported by Stenhammar et al.\cite{stenhammar2014phase}.

\begin{figure}[ht!]
    \centering 
    \includegraphics[width=\columnwidth]{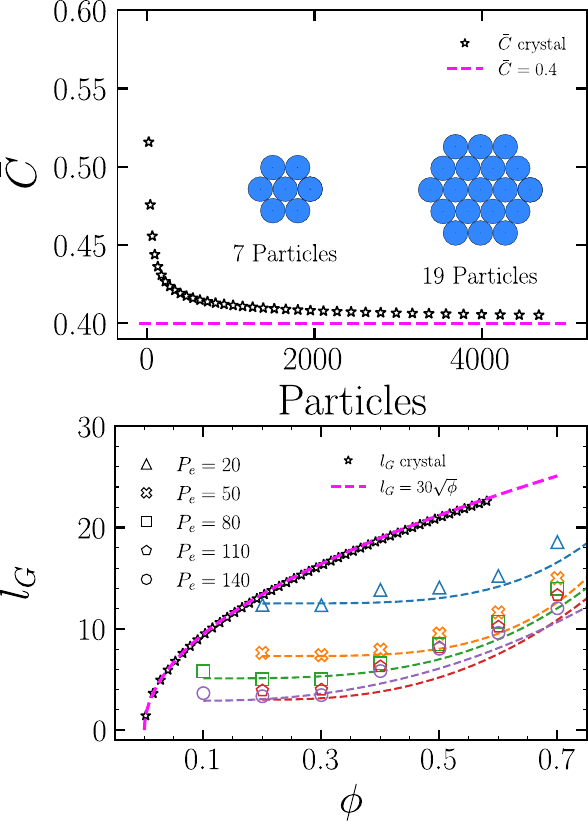}
    \caption{Top: Network average clustering coefficient $\bar{C}$ as a function of the particle number $N$. Following a crystal shape, $\bar{C}$ reaches an analytical saturation value $\bar{C}\sim 0.4$ as the cluster grows. Bottom: The asymptotic value of the network's average path length $l_G$ as a function of the packing fraction $\phi$. The star symbols represent the simulation's results for the case of a cluster growing in terms of $\phi$. The dashed magenta line represents the best fit, which follows a power-law behavior. In symbols, we present the asymptotic $l_G$ values measured from Fig.~\ref{fig:8} in terms of $\phi$ for different $\mbox{Pe}$. The dashed lines represent the best fits for each case using equation \eqref{lg}.}   
    \label{fig:9}%
\end{figure}

To make progress into understanding this behavior in Fig.~\ref{fig:9}, we show the average path length in the ideal case of perfect hexagonal clusters with $N$ particles, where we note that a power-law of the form $l_G = 30 \phi^{0.5}$ matches our numerical calculations shown in a magenta dashed line Fig~\ref{fig:9}. This relation means that while the cluster grows, the network is more connected; therefore, there are more paths connecting particles.

We propose the following phenomenological expression for the asymptotic values of the average path length in terms of $(\phi,\mbox{Pe})$ plotted in Fig~\ref{fig:9},

\begin{equation}
    l_G(\mbox{Pe},\phi) = \alpha +\beta \phi^{\zeta},
\end{equation}

where $\alpha$, $\beta$ and $\zeta$ are fitting parameters. We present the best-fit results in dashed lines of different colors in Fig.~\ref{fig:9} where we observe that for packing fractions $\phi<0.4$, the average path length is approximately constant at different P\'eclet numbers; this is represented in the fit parameter $\alpha$ which is inversely proportional to $\mbox{Pe}$.
For denser systems $\phi>0.4$, $l_G$ increases following a power-law with exponents between $\zeta\approx 2-4$, meanwhile $\beta\approx 20$ for all $\mbox{Pe}$. This behavior with increasing $\phi$ is a fingerprint of the cluster formation. When a solid phase appears in the steady state of the system, $l_G^{\text{solid}}$ contributes significantly to the network connection, suffering a drastic behavior when the phase separation emerges at large $\mbox{Pe}$.

\section{Conclusions}
In this work, we focused on studying the behavior of Active Brownian Particles (ABP) using complex networks. We examined the changes in the network structure as the packing fraction and Péclet number varied, particularly in the Motility-Induced Phase Separation (MIPS) region. We tracked the evolution of parameters such as the degree distribution, clustering coefficient, and average path length, revealing different network structures as the ABP behavior evolved. 

In the single-phase region, we conducted both numerical and theoretical analyses to explain the Gaussian distribution for random graphs. By analyzing the mean free path, we derived an expression for the average number of unique particle encounters within a given time window, highlighting the influence of packing fraction and Péclet number. Surprisingly, even in the single-phase region, we observed a critical packing fraction (\(\phi \geq \phi_c = \pi/(\pi + 2^{13/6}) \approx 0.41\)) that led to a significant change in network topology. Above this threshold, the mean free path became smaller than the contact radius (\(C_r\)) of the Weeks-Chandler-Andersen potential, indicating the pivotal role of packing fraction in network analysis.

We also identified a transient region where the degree distribution evolved over time, resulting in a hybrid distribution combining Gaussian and power law components. This region corresponded to the formation of small clusters without distinct separation. Our observations did not align with the analytical expression by Redner et al.~\cite{redner2013structure} but matched their numerical predictions and the critical cluster region reported by McDermott et al.~\cite{mcdermott2023characterizing} using machine learning analysis in MIPS. Additionally, the average path length and clustering coefficient stabilized in the hybrid region after \(t\geq 11 t_r\).

In the two-phase region, we found that the degree distribution followed a double power law distribution with two critical exponents, \(\gamma_1 \approx (1.5-3.7)\) and \(\gamma_2 \approx (12-17)\), for \(k>6\), representing solid and gas-like phases, respectively. We observed that \(\gamma_1<\gamma_2\), with \(\gamma_1\) reaching a stable state over time while \(\gamma_2\) fluctuated around an average value. Furthermore, we analyzed the average clustering coefficient in the presence of clustered particles. Due to the disk geometry of each ABP, we demonstrated that the system tended to approach values around \(\bar{C} = 0.4\) for large packing fractions. We discussed the behavior of the average path length (\(l_G\)), which increased with the packing fraction and decreased with the Péclet number and provided a heuristic model to describe its behavior across different regions.

Our research paves the way for future studies using complex networks to characterize phase transitions in various living organisms, from bacteria and ants to humans. We believe our findings will be relevant to ecological systems, transport phenomena, multi-agent-based models, and dynamic complex networks. Additionally, our work introduces possibilities for controlling and tracking clusterization in living and artificial active matter and their impact on disease transmission. The analytical predictions presented in this study can form the basis for investigating systems with complex interactions, including polarization, alignment, and taxis.

\bibliographystyle{unsrt}
\bibliography{ref}

\begin{thebibliography}{10}

\bibitem{romanczuk2012active}
Pawel Romanczuk, Markus B{\"a}r, Werner Ebeling, Benjamin Lindner, and Lutz Schimansky-Geier.
\newblock Active brownian particles: From individual to collective stochastic dynamics.
\newblock {\em The European Physical Journal Special Topics}, 202:1--162, 2012.

\bibitem{bechinger2016active}
Clemens Bechinger, Roberto Di~Leonardo, Hartmut L{\"o}wen, Charles Reichhardt, Giorgio Volpe, and Giovanni Volpe.
\newblock Active particles in complex and crowded environments.
\newblock {\em Reviews of Modern Physics}, 88(4):045006, 2016.

\bibitem{fily2012athermal}
Yaouen Fily and M~Cristina Marchetti.
\newblock Athermal phase separation of self-propelled particles with no alignment.
\newblock {\em Physical Review Letters}, 108(23):235702, 2012.

\bibitem{redner2013structure}
Gabriel~S Redner, Michael~F Hagan, and Aparna Baskaran.
\newblock Structure and dynamics of a phase-separating active colloidal fluid.
\newblock {\em Biophysical Journal}, 104(2):640a, 2013.

\bibitem{stenhammar2014phase}
Joakim Stenhammar, Davide Marenduzzo, Rosalind~J Allen, and Michael~E Cates.
\newblock Phase behaviour of active brownian particles: the role of dimensionality.
\newblock {\em Soft Matter}, 10(10):1489--1499, 2014.

\bibitem{theers2018clustering}
Mario Theers, Elmar Westphal, Kai Qi, Roland~G Winkler, and Gerhard Gompper.
\newblock Clustering of microswimmers: interplay of shape and hydrodynamics.
\newblock {\em Soft Matter}, 14(42):8590--8603, 2018.

\bibitem{van2019interrupted}
Marjolein~N Van Der~Linden, Lachlan~C Alexander, Dirk~GAL Aarts, and Olivier Dauchot.
\newblock Interrupted motility induced phase separation in aligning active colloids.
\newblock {\em Physical Review Letters}, 123(9):098001, 2019.

\bibitem{de2023sequential}
Pablo de~Castro, Felipe Urbina, Ariel Norambuena, and Francisca Guzm{\'a}n-Lastra.
\newblock Sequential epidemic-like spread between agglomerates of self-propelled agents in one dimension.
\newblock {\em Physical Review E}, 108(4):044104, 2023.

\bibitem{de2021active}
Pablo de~Castro, Saulo Diles, Rodrigo Soto, and Peter Sollich.
\newblock Active mixtures in a narrow channel: motility diversity changes cluster sizes.
\newblock {\em Soft Matter}, 17(8):2050--2061, 2021.

\bibitem{gompper20202020}
Gerhard Gompper, Roland~G Winkler, Thomas Speck, Alexandre Solon, Cesare Nardini, Fernando Peruani, Hartmut L{\"o}wen, Ramin Golestanian, U~Benjamin Kaupp, Luis Alvarez, et~al.
\newblock The 2020 motile active matter roadmap.
\newblock {\em Journal of Physics: Condensed Matter}, 32(19):193001, 2020.

\bibitem{caprini2022role}
Lorenzo Caprini, Rahul~Kumar Gupta, and Hartmut L{\"o}wen.
\newblock Role of rotational inertia for collective phenomena in active matter.
\newblock {\em Physical Chemistry Chemical Physics}, 24(40):24910--24916, 2022.

\bibitem{zottl2023modeling}
Andreas Z{\"o}ttl and Holger Stark.
\newblock Modeling active colloids: From active brownian particles to hydrodynamic and chemical fields.
\newblock {\em Annual Review of Condensed Matter Physics}, 14:109--127, 2023.

\bibitem{tan2022odd}
Tzer~Han Tan, Alexander Mietke, Junang Li, Yuchao Chen, Hugh Higinbotham, Peter~J Foster, Shreyas Gokhale, J{\"o}rn Dunkel, and Nikta Fakhri.
\newblock Odd dynamics of living chiral crystals.
\newblock {\em Nature}, 607(7918):287--293, 2022.

\bibitem{petroff2015fast}
Alexander~P Petroff, Xiao-Lun Wu, and Albert Libchaber.
\newblock Fast-moving bacteria self-organize into active two-dimensional crystals of rotating cells.
\newblock {\em Physical Review Letters}, 114(15):158102, 2015.

\bibitem{ridgway2023motility}
Wesley~JM Ridgway, Mohit~P Dalwadi, Philip Pearce, and S~Jonathan Chapman.
\newblock Motility-induced phase separation mediated by bacterial quorum sensing.
\newblock {\em bioRxiv}, pages 2023--04, 2023.

\bibitem{norambuena2020understanding}
Ariel Norambuena, Felipe~J Valencia, and Francisca Guzm{\'a}n-Lastra.
\newblock Understanding contagion dynamics through microscopic processes in active brownian particles.
\newblock {\em Scientific Reports}, 10(1):20845, 2020.

\bibitem{forgacs2022using}
P~Forg{\'a}cs, A~Lib{\'a}l, C~Reichhardt, N~Hengartner, and CJO Reichhardt.
\newblock Using active matter to introduce spatial heterogeneity to the susceptible infected recovered model of epidemic spreading.
\newblock {\em Scientific Reports}, 12(1):11229, 2022.

\bibitem{richardson2015beyond}
Thomas~O Richardson and Thomas~E Gorochowski.
\newblock Beyond contact-based transmission networks: the role of spatial coincidence.
\newblock {\em Journal of The Royal Society Interface}, 12(111):20150705, 2015.

\bibitem{gorochowski2017behaviour}
Thomas~E Gorochowski and Thomas~O Richardson.
\newblock How behaviour and the environment influence transmission in mobile groups.
\newblock {\em Temporal Network Epidemiology}, pages 17--42, 2017.

\bibitem{zhong2023burstiness}
Wei Zhong, Youjin Deng, and Daxing Xiong.
\newblock Burstiness and information spreading in active particle systems.
\newblock {\em Soft Matter}, 19(16):2962--2969, 2023.

\bibitem{bhaskar2021topological}
Dhananjay Bhaskar, William~Y Zhang, and Ian~Y Wong.
\newblock Topological data analysis of collective and individual epithelial cells using persistent homology of loops.
\newblock {\em Soft Matter}, 17(17):4653--4664, 2021.

\bibitem{mcdermott2023characterizing}
D~McDermott, C~Reichhardt, and CJO Reichhardt.
\newblock Characterizing different motility-induced regimes in active matter with machine learning and noise.
\newblock {\em Physical Review E}, 108(6):064613, 2023.

\bibitem{newman2001}
M.~E.~J. Newman.
\newblock \textsl{Scientific collaboration networks. II. Shortest paths, weighted networks, and centrality.}
\newblock {\em \textit{Physical Review Journals.}}, \textbf{64}, 2001.

\bibitem{Kertesz}
J.~Kertész, J.~Török, Y.~Murase, and K.~Jo, HH.and~Kaski.
\newblock {\em Modeling the Complex Network of Social Interactions. In: Rudas, T., Péli, G. (eds) Pathways Between Social Science and Computational Social Science.}
\newblock Computational Social Sciences.Springer, Cham., https://doi.org/10.1007/978-3-030-54936-7-1, 2021.

\bibitem{gonzalez2008understanding}
Marta~C Gonzalez, Cesar~A Hidalgo, and Albert-Laszlo Barabasi.
\newblock Understanding individual human mobility patterns.
\newblock {\em Nature}, 453(7196):779--782, 2008.

\bibitem{abe2006}
S.~Abe and N.~Suzuki.
\newblock \textsl{Complex-network description of seismicity.}
\newblock {\em \textit{Nonlinear Proc. Geophys}}, \textbf{13}:145--150, 2006.

\bibitem{telesca2012}
L.~Telesca and M.~Lovallo.
\newblock \textsl{Analysis of seismic sequences by using the method of visibility graph. }.
\newblock {\em \textit{Europhysics Letters.}}, \textbf{97}, 2012.

\bibitem{pasten2016}
D.Pastén, F.~Torres, B.~Toledo, V.~Mu\ noz, J.~Rogan, and J.~A. Valdivia.
\newblock \textsl{Time-Based Network Analysis Before and After the M w 8.3 Illapel Earthquake 2015 Chile.}
\newblock {\em \textit{Pure and Applied Geophysics.}}, \textbf{173}:2267--2275, 2016.

\bibitem{pasten2018}
D.~Past\'en, Z.~Czechowski, and B.~Toledo.
\newblock \textsl{Time series analysis in earthquake complex networks.}
\newblock {\em \textit{Chaos: An Interdisciplinary Journal of Nonlinear Science.}}, \textbf{28}, 2018.

\bibitem{suyal2014visibility}
Vinita Suyal, Awadhesh Prasad, and Harinder~P Singh.
\newblock Visibility-graph analysis of the solar wind velocity.
\newblock {\em Solar Physics}, 289(1):379--389, 2014.

\bibitem{mohammadi2021complex}
Z~Mohammadi, N~Alipour, H~Safari, and Farhad Zamani.
\newblock Complex network for solar protons and correlations with flares.
\newblock {\em Journal of Geophysical Research: Space Physics}, 126(7):e2020JA028868, 2021.

\bibitem{thiery}
J.~Thiery and J.~Sleeman.
\newblock Complex networks orchestrate epithelial–mesenchymal transitions.
\newblock {\em Nat. Rev. Mol. Cell. Biol.}, 7:131--142, 2006.

\bibitem{barabasi2011}
A.L. Barabási, N.~Gulbahce, and J.~Loscalzo.
\newblock Network medicine: A network-based approach to human disease.
\newblock {\em Nat. Rev. Genet.}, 12:56--68, 2011.

\bibitem{scabini}
L.F.S Scabini, L.C. Ribas, M.B. Neiva, A.G.B Junior, A.J.F Farfán, and O.M. Bruno.
\newblock Social interaction layers in complex networks for the dynamical epidemic modeling of covid-19 in brazil.
\newblock {\em Physica A}, 564(125498):ISSN 0378--4371, 2021.

\bibitem{riley2007large}
Steven Riley.
\newblock Large-scale spatial-transmission models of infectious disease.
\newblock {\em Science}, 316(5829):1298--1301, 2007.

\bibitem{stockmaier2021infectious}
Sebastian Stockmaier, Nathalie Stroeymeyt, Eric~C Shattuck, Dana~M Hawley, Lauren~Ancel Meyers, and Daniel~I Bolnick.
\newblock Infectious diseases and social distancing in nature.
\newblock {\em Science}, 371(6533):eabc8881, 2021.

\bibitem{newman2003}
M.~E.~J. Newman.
\newblock \textsl{The Structure and Function of Complex Networks.}
\newblock {\em \textit{Society for Industrial and Applied Mathematics.}}, \textbf{45}:167--256, 2003.

\bibitem{albert2002statistical}
R\'eka Albert and Albert-L\'aszl\'o Barab\'asi.
\newblock Statistical mechanics of complex networks.
\newblock {\em Rev. Mod. Phys.}, 74:47--97, Jan 2002.

\bibitem{erdos1960}
Erdös P. and A.~Rényi.
\newblock On the evolution of random graphs.
\newblock {\em Publ. math. inst. hung. acad. sci}, 5(1):17--60, 1960.

\bibitem{barabasi1999}
Albert-László Barabási and Réka Albert.
\newblock Emergence of scaling in random networks.
\newblock {\em Science}, 286(5439):509--512, 1999.

\bibitem{watts1998}
D.~J. Watts and S.~H. Strogatz.
\newblock Collective dynamics of 'small-world' networks.
\newblock {\em Nature}, 393(6684):440--442, 1998.

\bibitem{cates2015motility}
Michael~E Cates and Julien Tailleur.
\newblock Motility-induced phase separation.
\newblock {\em Annu. Rev. Condens. Matter Phys.}, 6(1):219--244, 2015.

\bibitem{buttinoni2013dynamical}
Ivo Buttinoni, Julian Bialk{\'e}, Felix K{\"u}mmel, Hartmut L{\"o}wen, Clemens Bechinger, and Thomas Speck.
\newblock Dynamical clustering and phase separation in suspensions of self-propelled colloidal particles.
\newblock {\em Physical Review Letters}, 110(23):238301, 2013.

\bibitem{su2023motility}
Jie Su, Mengkai Feng, Yunfei Du, Huijun Jiang, and Zhonghuai Hou.
\newblock Motility-induced phase separation is reentrant.
\newblock {\em Communications Physics}, 6(1):58, 2023.

\bibitem{Erdos}
P.~Erdős and A.~Rényi.
\newblock On random graphs. i.
\newblock {\em Publicationes Mathematicae}, 6 (3-4):290--297, 1959.

\bibitem{Newman_2000}
M.~E.~J. Newman, C.~Moore, and D.~J. Watts.
\newblock Mean-field solution of the small-world network model.
\newblock {\em Phys. Rev. Lett.}, 84:3201--3204, Apr 2000.

\bibitem{newman_watts1999}
M.~E.~J. Newman and D.~J. Watts.
\newblock Scaling and percolation in the small-world network model.
\newblock {\em Phys. Rev. E}, 60:7332--7342, Dec 1999.

\bibitem{soto2024kinetic}
Rodrigo Soto, Martin Pinto, and Ricardo Brito.
\newblock Kinetic theory of motility induced phase separation for active brownian particles.
\newblock 2024.

\bibitem{Paik_2014}
Steve~T. Paik.
\newblock Is the mean free path the mean of a distribution?
\newblock {\em American Journal of Physics}, 82(6):602–608, June 2014.

\bibitem{Li2004}
W.~Li and X.~Cai.
\newblock Statistical analysis of airport network of china.
\newblock {\em Phys. Rev. E}, 69:046106, Apr 2004.

\bibitem{Li-Ping:1393}
CHI Li-Ping, WANG Ru, SU~Hang, XU~Xin-Ping, ZHAO Jin-Song, LI~Wei, and CAI Xu.
\newblock Structural properties of us flight network.
\newblock {\em Chinese Physics Letters}, 20(8):1393--1396, 2003.

\bibitem{Han_2011}
D.~D. Han, J.~H. Qian, and Y.~G. Ma.
\newblock Emergence of double scaling law in complex systems.
\newblock {\em Europhysics Letters}, 94(2):28006, apr 2011.

\bibitem{ZHANG20103922}
Jun Zhang, Xian-Bin Cao, Wen-Bo Du, and Kai-Quan Cai.
\newblock Evolution of chinese airport network.
\newblock {\em Physica A: Statistical Mechanics and its Applications}, 389(18):3922--3931, 2010.

\bibitem{Dorogovtsev2001}
S.~N. Dorogovtsev and J.~F.~F. Mendes.
\newblock Language as an evolving word web.
\newblock {\em Proc. R. Soc. Lond. B.}, 268:2603–2606, 2001.

\bibitem{PINTO20144019}
Carla~M.A. Pinto, A.~Mendes Lopes, and J.A. {Tenreiro Machado}.
\newblock Double power laws, fractals and self-similarity.
\newblock {\em Applied Mathematical Modelling}, 38(15):4019--4026, 2014.

\end{thebibliography}

\section*{Acknowledgments}
The authors are grateful for the helpful feedback given by Pablo de Castro. I.S and F.G.-L. have received support from the ANID – Millennium Science Initiative Program – NCN19 170, Chile. F.G.-L. was supported by Fondecyt Iniciación No.\ 11220683. A.N. acknowledges the financial support from the project Fondecyt Iniciaci\'on No. 11220266.

%\bibliography{ref}% Produces the bibliography via BibTeX.

\end{document}